\documentclass[aps,prl,twocolumn,superscriptaddress]{revtex4}
\usepackage{here}
\usepackage{graphicx}
\usepackage[usenames]{color}
\newcommand{\tn}{$T_N$}

\newcommand{\oo}{ $(1/2,\ 0,\ 0)$}
\newcommand{\mo}{ $(0,\ 1/2,\ 0)$}

\begin{document}
\title{Electronic phase transitions in 
Pr$_{0.5}$Ca$_{0.5}$MnO$_3$ epitaxial thin films 
revealed by resonant soft x-ray scattering}

\author{H.~Wadati}
\email{wadati@ap.t.u-tokyo.ac.jp}
\homepage{http://www.geocities.jp/qxbqd097/index2.htm}
\affiliation{Department of Physics and Astronomy, 
University of British Columbia, 
Vancouver, British Columbia V6T 1Z1, Canada}
\affiliation{Department of Applied Physics and Quantum-Phase 
Electronics Center (QPEC), University of Tokyo, Hongo, 
Tokyo 113-8656, Japan}

\author{J. Geck}
\affiliation{IFW Dresden, P.O. Box 270116, D-01171 Dresden, Germany}

\author{E. Schierle}
\affiliation{Helmholtz-Zentrum Berlin 
f\"{u}r Materialien und Energie 
Albert-Einstein-Stra\ss e 15, 
12489 Berlin, Germany}

\author{R. Sutarto} 
\affiliation{Department of Physics and Astronomy, 
University of British Columbia, 
Vancouver, British Columbia V6T 1Z1, Canada}

\author{F. He} 
\affiliation{Canadian Light Source, University of 
Saskatchewan, Saskatoon, Saskatchewan S7N 0X4, Canada}

\author{D. G. Hawthorn}
\affiliation{Department of Physics and Astronomy, 
University of Waterloo, Waterloo, Ontario N2L 3G1, Canada}

\author{M.~Nakamura}
\affiliation{Cross-Correlated Materials Research Group (CMRG), 
ASI, RIKEN, Wako 351-0198, Japan}

\author{M. Kawasaki}
\affiliation{Department of Applied Physics and Quantum-Phase 
Electronics Center (QPEC), University of Tokyo, Hongo, 
Tokyo 113-8656, Japan} 
\affiliation{Cross-Correlated Materials Research Group (CMRG), 
ASI, RIKEN, Wako 351-0198, Japan}

\author{Y. Tokura}
\affiliation{Department of Applied Physics and Quantum-Phase 
Electronics Center (QPEC), University of Tokyo, Hongo, 
Tokyo 113-8656, Japan} 
\affiliation{Cross-Correlated Materials Research Group (CMRG), 
ASI, RIKEN, Wako 351-0198, Japan}

\author{G.~A.~Sawatzky}
\affiliation{Department of Physics and Astronomy, 
University of British Columbia, 
Vancouver, British Columbia V6T 1Z1, Canada}


\date{\today}
\begin{abstract}
We report the study of magnetic and orbital order 
in Pr$_{0.5}$Ca$_{0.5}$MnO$_3$ epitaxial thin films 
grown on 
(LaAlO$_3$)$_{0.3}$-(SrAl$_{0.5}$Ta$_{0.5}$O$_3$)$_{0.7}$ (LSAT) (011)$_c$. 
In a new experimental approach, the polarization and 
energy dependence of resonant soft x-ray scattering are used to 
reveal significant modifications of the magnetic order in the film 
as compared to the bulk, namely 
(i) a different magnetic ordering wave vector, 
(ii) a different magnetic easy axis and (iii) an additional magnetic 
reordering transition at low temperatures. 
These observations indicate a strong impact of the epitaxial 
strain on the spin order, which is mediated by the orbital 
degrees of freedom and which provides a promising route 
to tune the magnetic properties of manganite films. 
Our results further demonstrate that resonant soft x-ray scattering 
is a very suitable technique to study the magnetism in thin films, 
to which neutron scattering cannot easily be applied due to the small sample volume. 
\end{abstract}
\maketitle
Hole-doped perovskite manganites 
$R_{1-x}A_x$MnO$_3$, where $R$ is a rare-earth 
($R=$ La, Nd, Pr) and $A$ is an alkaline-earth atom 
($A=$ Sr, Ba, Ca) have attracted much attention 
because they exhibit remarkable physical properties 
such as colossal magnetoresistance and 
complex electronic ordering phenomena 
\cite{rev, RamirezMn,orbital,RaoMn,PrellierMn,DagottoMn,
Hungry,TokuraMn}. 
For the latter, the half-doped manganites provide 
a particular prominent and extensively studied example, 
namely the so-called CE-phase \cite{JirakPCMO}. 
This phase is commonly discussed in terms of cooperative spin, 
charge and orbital order, where ferromagnetic zig-zag chains are formed, 
which are coupled antiferromagnetically to each other. Notwithstanding 
intensive research, however, the detailed microscopic structure of the
electronic order remains to be fully understood. In particular, 
the question whether charge ordering really exists and, in case it does, 
whether it is centered on oxygen or manganese is currently 
discussed vigorously in the literature \cite{Zenerp, Khom}. 

Potential applications of manganites heavily rely 
on their physical properties in the form of thin films grown 
on a substrate. Epitaxial thin films are usually 
strained to some degree and the interfaces may introduce 
modifications of the properties. 
It is therefore very important to investigate and understand 
the electronic modifications in doped manganite films, 
which can be dramatic. 
For example, it was recently shown that epitaxial strain 
effects can control charge ordering (CO) in thin films of 
Mn-oxides \cite{NSMONakamura,okumura}. 
A transition between CO and ferromagnetic metallic states 
was observed in Nd$_{0.5}$Sr$_{0.5}$MnO$_3$ films 
on SrTiO$_3$ (011)$_c$ substrates, whereas Nd$_{0.5}$Sr$_{0.5}$MnO$_3$ 
films on SrTiO$_3$ (001)$_c$ substrates exhibit only insulating behavior 
at all temperatures \cite{NSMONakamura}. 
Here the substrate orientation is given in 
the standard cubic notation. 
Also in Pr$_{0.5}$Ca$_{0.5}$MnO$_3$ thin films, 
epitaxial strain strongly affects the electronic properties. 
Previous studies have shown that 
Pr$_{0.5}$Ca$_{0.5}$MnO$_3$ films grown epitaxially on 
(LaAlO$_3$)$_{0.3}$-(SrAl$_{0.5}$Ta$_{0.5}$O$_3$)$_{0.7}$ (LSAT)
(011)$_c$ substrate exhibit a CO transition around 220 K, similar to bulk samples, 
while Pr$_{0.5}$Ca$_{0.5}$MnO$_3$ films on LSAT (001)$_c$ 
substrates has a much higher CO transition 
temperature around 300K \cite{okumura}. 

Here we present a resonant soft x-ray scattering (RSXS) study of the 
electronic order in Pr$_{0.5}$Ca$_{0.5}$MnO$_3$ 
thin films grown epitaxially on LSAT (011), which implies that 
the microscopic magnetic order in films can differ significantly 
from that of the corresponding bulk materials. 
Our results indicate that epitaxial strain couples 
to the spin order via the orbital degrees of freedom, 
which provides a unique way to tune the magnetic properties of doped manganite films. 

Pr$_{0.5}$Ca$_{0.5}$MnO$_3$ thin films were grown on LSAT (011)$_c$ substrates 
by pulsed laser deposition. 
Details of the fabrication and characterization 
of the thin films were described elsewhere \cite{okumura}.  
RSXS experiments at the Mn 2p edge were performed at the BESSY
undulator beamline UE 46-PGM and 
10ID-2 (REIXS) 
of the Canadian Light Source \cite{David}. 
Scattering spectra were measured using horizontally ($\pi$) or vertically 
polarized ($\sigma$) light. 
We will refer to the resonant intensity measured in the 
$\pi\rightarrow\sigma^{\prime}$, $\pi^{\prime}$ and 
$\sigma\rightarrow\sigma^{\prime}$, $\pi^{\prime}$ channels as $I_\pi$ 
and $I_\sigma$, respectively. 
The pressure during measurements were below $5\times 10^{-9}$ Torr, 
and the temperature was varied 
between room temperature (RT) and 25 K. 
We also performed x-ray absorption spectroscopy 
(XAS) measurements 
in the total-electron-yield (TEY) mode. 

\begin{figure}
\begin{center}
\includegraphics[width=8cm]{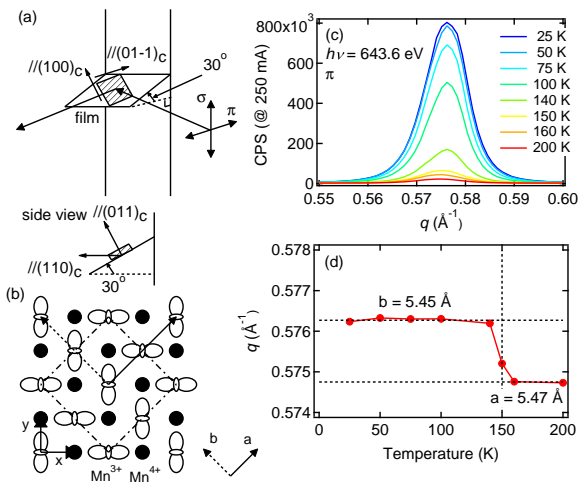}
\caption{(Color online) 
Temperature dependence observed at the $(1/2, 0, 0)_{D1}$ position (c). 
The experimental geometry is shown in panel (a). 
Panel (b) shows the schematic drawing of the charge-orbital ordering. 
The arrow marks the directions for the orbital ordering and 
the dashed arrow for the spin ordering. 
Panel (d) shows the peak positions as a function of temperature.}
\label{fig1}
\end{center}
\end{figure}

In the following, the $(HKL)$-indexes for the film reflections and 
directions refer to the orthorhombic unit cell of the 
PCMO-film\,\cite{okumura}. At this point it is important to realize 
that the PCMO-films contain structural domains with interchanged $a$-
and $b$-axes -- so-called twin domains.  
We will refer to these twin domains as $D1$ and $D2$, 
respectively, and designate the reflections and directions of these 
domains with a corresponding index. Using this convention, 
the orthorhombic $(100)_{D1}$ and $(011)_{D1}$ directions 
were parallel to the scattering plane as shown 
in Fig.~\ref{fig1} (a), which corresponds to 
$(010)_{D2}$ and $(101)_{D2}$ of the other twin domain. 
In a scattering experiment, the diffraction patterns 
of $D\,1$ and $D\,2$ are superimposed. 
This means that close to the position of 
$(H,0,0)_{D\,1}$, the $(0,H,0)_{D\,2}$ reflection can be observed, 
if it occurs as well. However, because  $a\neq b$, these two 
reflections posses slightly different scattering angles. 
This orthorhombic peak splitting enables 
to identify the $(H,0,0)_{D\,1}$ and $(0,H,0)_{D\,2}$ reflections from 
the different domains.

Figure \ref{fig1} (c) shows the temperature dependence 
of the resonant intensity measured around the  $(1/2, 0, 0)_{D1}$ 
position with incoming $\pi$-polarization. The photon energy was set to
643.6 eV, which corresponds to 
the Mn $2p$ absorption peak. A clear peak appears around 210 K and
strongly gains 
intensity with decreasing temperature. This result is 
in good agreement with the charge/orbital-ordering transition 
temperature $T_{\mathrm{CO/OO}}= 220$ K reported in 
Ref.~\cite{okumura}. 
The peak widths show almost no change with 
temperature, demonstrating that 
the ordering is always long-ranged below $T_{CO/OO}$ 
and that the domain sizes do 
not exhibit a significant temperature dependence. 

The position of the superlattice peak as a function of temperature 
is shown in Fig.~\ref{fig1} (d).  
A clear shift of the peak position is observed at the magnetic 
ordering temperature $T_N$ = 150 K, even though the lattice 
parameters do not change significantly at this 
temperature \cite{okumura}. Above $T_N$ and below $T_{CO/OO}$, 
the peak position corresponds exactly to the 
$(1/2,0,0)_{D1}$ reflection, identifying it as the superlattice 
reflection due to the orbital order in $D1$. Note that the orbital order 
causes a doubling of the orthorhombic $a$-axis only, i.e., 
the $(0,1/2,0)_{D2}$ does not occur in the orbital ordered phase. 
However, with the onset of magnetic order below $T_N$, 
the peak position jumps to $(0,1/2,0)_{D2}$, implying that 
the observed intensity is now due to the magnetic order 
in twin domain $D\,2$. As we will discuss further below, 
the orbital scattering of $D1$ still exists below $T_N$, 
but its intensity is much smaller 
than the magnetic scattering of $D2$.

\begin{figure}
\begin{center}
\includegraphics[width=9cm]{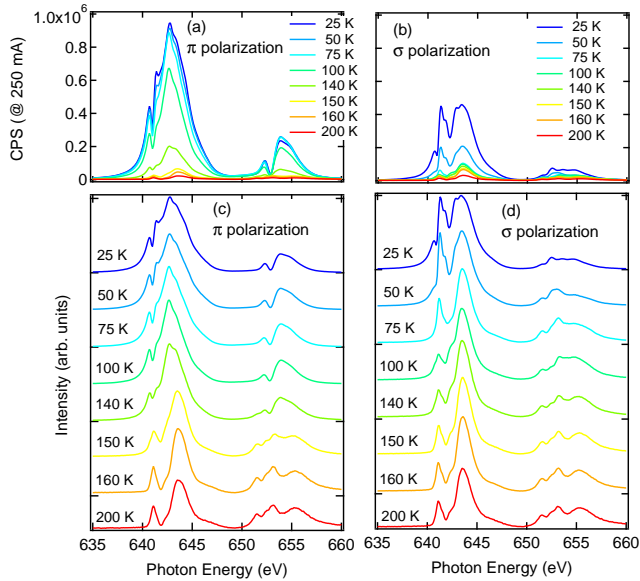}
\caption{(Color online) 
Photon-energy dependence of the (1/2, 0, 0)/(0, 1/2, 0) peak intensity at Mn $2p$ 
edges at various temperatures using $\pi$ polarization (a) and $\sigma$ 
polarization (b). To compare lineshapes, 
all data in the panel (a) and (b) have been normalized to each 
spectrum area, shown at the respective panel 
(c) and (d) as a function of photon energies.}
\label{fig2}
\end{center}
\end{figure}

Figure \ref{fig2} shows the photon-energy dependence of 
the intensity observed at the 
$(1/2, 0, 0)_{D1}$ position 
across the Mn $2p$ edges at various temperatures 
using $\pi$ and $\sigma$ polarizations. 
To facilitate a comparison of the lineshape, panels (c) and (d) 
show the same data as given in (a) and (b), 
but this time normalized to the 
area. As can be observed in Fig.~\ref{fig2}, 
$I_\pi$ and $I_\sigma$ are of very similar 
magnitude and exhibit the same lineshapes for 150 K $<T<$ 200 K, 
i.e., $I_\pi \simeq I_\sigma$ for the orbital $(1/2,0,0)_{D1}$ peak. 
The situation clearly changes at 150\,K upon cooling: while $I_\pi$ 
shows a strong increases by a factor of $\sim 10$ 
accompanied by a clear lineshape change, $I_\sigma$ remains almost 
unaltered. Furthermore, whereas the peak in $I_\pi$ 
apparently shifts in position, as described above 
(cf. Fig.\,\ref{fig1}), the peak in $I_\sigma$ does not move at $T_N$. 
From this we conclude that the additional magnetic 
scattering below $T_N$, 
which corresponds to $(0,1/2,0)_{D2}$ of $D2$, is almost entirely 
restricted to $I_\pi$, whereas $I_\sigma$ is due to the orbital
$(1/2,0,0)_{D1}$ peak of $D1$. 

Surprisingly, we observe another dramatic change of the RSXS linshapes 
at $T_2=75$ K well below $T_N$. This time, $I_\sigma$ displays clear
changes of lineshape and intensity, as shown 
in Figs.~\ref{fig2} (b) and (d). 
The variation of $I_\sigma$ across $T_2$ signals a third phase
transition, which has not been reported earlier and is absent 
in the bulk material. 

\begin{figure}
\begin{center}
\includegraphics[width=9cm]{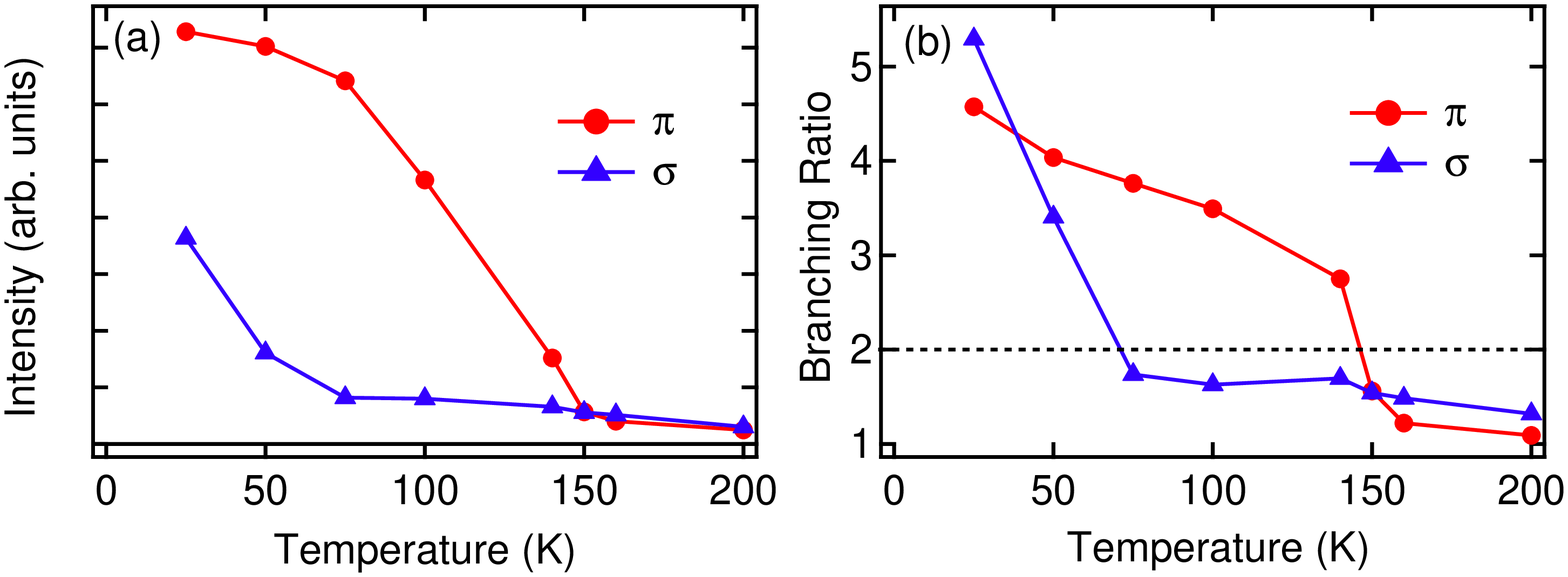}
\caption{(Color online) The integrated intensities of the 
(1/2, 0, 0)/(0, 1/2, 0) 
peak over both the Mn $2p_{3/2}$ and Mn $2p_{1/2}$ regions (a) and 
the branching ratio (b), where a statistical value of 2 is 
also indicated by the dash line.}
\label{fig4}
\end{center}
\end{figure}

The phase transitions at $T_N$ and $T_2$ are also clearly revealed 
by the data presented in Fig.~\ref{fig4} (a), 
which shows $I_\pi$ and $I_\sigma$ integrated over 
the Mn $2p_{3/2}$ and $2p_{1/2}$ regions. 
In these three panels, one can see a large increase of the total 
scattering intensity by a factor of more than 10 in $I_\pi$ 
around 150 K. This demonstrates the sensitivity of the RSXS-intensity 
in going from orbital to orbital plus spin order. $I_\sigma$ 
increases strongly around $T_2=75$ K. 

The changes of the lineshapes at $T_N$ an $T_2$ 
become even more apparent by plotting the so-called branching 
ratio in Fig.~\ref{fig4} (b). 
Branching ratios are defined by the intensity contributions 
from the Mn $2p_{3/2}$ region divided 
by that from the Mn $2p_{1/2}$ region. 
Since the Mn $2p_{3/2}$ and $2p_{1/2}$ states have 4 and 2 states, 
respectively, the statistical values of branching ratios are 2, 
but this value changes due to the local crystal fields, 
orbital ordering as well as the spin orientation in the 
ordered system \cite{Bratio}. 
The branching ratios changes strongly around 150 K 
for a $\pi$ polarization, 
and similarly around 75 K for a $\sigma$ polarization. 

In order to gain a better understanding of the experimental results, 
we have investigated the resonant structure factor $F$ at the Mn $2p$ edges. 
Referring to the standard model of the CE-phase, we describe 
the so-called Mn$^{3+}$-sites within a local $D_{4h}$-symmetry, 
which corresponds to the occupied orbital, 
while the local symmetry of the so-called Mn$^{4+}$-sites is taken to 
be $O_h$ (cf. Fig.~\ref{fig1}). Note, that the present model refers 
to local symmetries only and does not make any assumptions 
regarding the charge disproportionation in the CE-phase. 

The resonant scattering length of a given site $n$ can be expressed as 
$f_n=\mathbf{e}_f^+.\hat f_n(\hbar\omega,\mathbf{\hat s}).\mathbf{e}_i$, 
with $\hbar \omega$ the photon energy, $\mathbf{\hat s}=(s_x,s_y,s_z)$ 
the direction of the local spin, and $\mathbf{e}_i$ ($\mathbf{e}_f$) 
the polarization of the incident (scattered) beam. 
For the $\hat f_n$ in $O_h$ and $D_{4h}$ we use formulae given in 
Ref.~\cite{Maurits}, which express the ($3\times3$)-matrixes $\hat f_n$ 
in terms of a few fundamental spectra $f^{(k)}_{n,j}(\hbar \omega)$ 
with $k=0$, 1, 2, 3. At the Mn $2p_{3/2,1/2}$ resonance, the scattering is 
completely dominated by the Mn-subblattice. We therefore calculate 
the resonant structure factor matrix 
$\hat F(\mathbf{Q})=\sum_n \hat f_n
\exp(i \mathbf{Q}.\mathbf{r}_n)=\sum_{k=0}^{3}\hat F^{(k)}(\mathbf{Q})$ 
by summing over the 16 Mn-sites of the CE-supercell. Here, 
$\hat F^{(k)}$ only contains fundamental spectra 
with fixed $k$.  The scattering amplitude and intensity 
for given $\mathbf{e}_i$ and $\mathbf{e}_f$ can then be calculated 
as $A_{if}=\mathbf{e}_i^+.\hat F.\mathbf{e}_f$ and 
$I_{if}\propto |A_{if}|^2$, respectively. 

For $T_{\mathrm{CO/OO}}>T>$ \tn, i.e., for orbital order without 
spin order, 
the calculated scattering amplitudes of the orbital \oo\/ reflection are 
$A_{\sigma\sigma'}^{orb}=A_{\pi\pi'}^{orb}=0$ and 
$A_{\sigma\pi'}^{orb}=A_{\pi\sigma'}^{orb}\propto
(f^{(0)}_{a1_g,x}-f^{(0)}_{a1_g,y})$, 
where $f^{(0)}_{a1_g,x}$ and $f^{(0)}_{a1_g,y}$ are fundamental 
spectra with $k=0$. These spectra describe the anisotropy with 
respect to the local $C_4$-axis in $D_{4h}$, i.e., the anisotropy 
for polarizations along or perpendicular to the occupied orbital 
at the Mn$^{3+}$-sites. The calculation gives $\sigma\pi'$- and
$\pi\sigma'$-scattering 
only and yields $I_\pi=I_\sigma$, which is identical to previous 
results and in good agreement with the present experiment. 

A more interesting situation arises when orbital and magnetic 
order coexist ($T<$ \tn ). In this case, the structure factor of 
the orbital order reflection is of the form 
$\hat F_{orb}=\hat F^{(0)}_{orb}+\hat F^{(2)}_{orb}$ 
($\mathbf{\hat s}$) ($\mathbf{\hat s}$: easy axis). 
Somewhat surprising, the above equation shows that the spin 
order can change the resonant scattering at the orbital ordering peak, 
even though it  does not correspond to the magnetic ordering wave vector. 
These additional contributions are related to the magnetic linear 
dichroism in XAS and are even in $\mathbf{\hat s}$ \cite{Maurits}.
On first sight, the spin-dependent $F^{(2)}_{orb}$ might explain 
the observed dramatic changes of $I_\pi$ below \tn. However, as discussed above, 
the shift in position accompanying the increase of $I_\pi$ implies 
that the additional intensity has to be attributed to the magnetic 
(0, 1/2, 0) superlattice peak. 
The structure factor of this magnetic reflection is given by 
$\hat F_{mag}=\hat F^{(1)}_{mag}(\mathbf{\hat s})
+\hat F^{(3)}_{mag}(\mathbf{\hat s})\, .$ 
These terms are related to the circular magnetic dichroism in
XAS\,\cite{Maurits}, 
i.e., the \mo\/ reflection is given by terms odd in $\mathbf{\hat s}$ 
as it should be. 

Comparing $F_{orb}$ and $F_{mag}$ to the experimental 
data a number of important conclusions can be drawn: 
Since the orbital scattering in  $I_\sigma$ due to the 
$(1/2,0,0)_{D1}$ peak does not change much for $T_2<T<T_{CO/OO}$, 
it is still dominated by same fundamental spectra 
$f^{(0)}_{a1_g,x}$ and  $f^{(0)}_{a1_g,y}$ at these temperatures. 
In contrast to this, the lineshape in $I_\pi$ does change 
dramatically with the onset of magnetic order, due to the 
strong additional magnetic intensity stemming 
from the $(0,1/2,0)_{D2}$ peak. The changed lineshape is easily 
explained qualitatively by the different 
fundamental spectra, which enter $\hat F_{orb}$ and $\hat F_{mag}$. 

Furthermore, the experimental results show that for \tn$>T>T_2$ 
the magnetic intensity is almost exclusively confined to $I_\pi$, 
while there is almost no magnetic scattering in $I_\sigma$. 
This very specific feature of the magnetic scattering provides information about the spin directions in this temperature range: 
We first note that, since $|f^{(k=3)}_{n,j}|<<|f^{(k=1)}_{n,j}|$ (cf. Ref.\,\cite{Maurits}), we can take the approximation 
$\hat F = \hat F^{(1)}$, which yields 
\begin{eqnarray*}
\sqrt{3}\, A^{mag}_{\sigma\pi'}&=& f^{(1)}_{a_{2u}} \left(\left(2+3 \sqrt{2}\right) s_x+\left(3 \sqrt{2}-2\right) s_y\right)\\
&+& f^{(1)}_{e_{u}}\ \left(\left(2+3 \sqrt{2}\right) s_x+\left(3 \sqrt{2}-2\right) s_y +4 s_z\right)
\end{eqnarray*}
$s_{x,y,z}$ refer to the $x,y,z$-axes indicated in Fig.\,\ref{fig1}. No 
magnetic scattering in $I_\sigma$ means that $A^{mag}_{\sigma\pi'}=0$ 
at all photon energies. Since $f^{(1)}_{a_{2u}}$ and $f^{(1)}_{e_{u}}$ 
are linear independent, their coefficients must vanish in this case, 
which implies that $s_z=0$ and $s_y/s_x=(2+3\sqrt{2})/(2-3\sqrt{2})$ 
and corresponds to an easy axis close to 
$\mathbf{\hat s}_0=(\cos \phi,\sin \phi,0)$ with $\phi=110^{\circ}$.  
Note that the spin orientation determines how the different fundamental 
spectra contribute to the total intensity. We therefore conclude 
that the dramatic changes of $I_\sigma$ below $T_2$ are due 
to a spin reorientation, which results in $A_{\sigma\pi'}\neq0$. 
This additional spin reorientation clearly indices that the epitaxial 
strain changes the magnetic anisotropy of PCMO. 

Finally we note that for spins close to $\mathbf{\hat s}_0$ 
the term $F^{(2)}_{orb}$ should result in additional contributions 
to $I_\sigma$ as well. 
But since the lineshape 
of $I_\sigma$ remains largely unaltered across \tn, the data 
indicates that these contributions are significantly smaller 
than the ones given by $F^{(0)}_{orb}$.  

In summary, we performed RSXS studies 
of the orbital and magnetic order in Pr$_{0.5}$Ca$_{0.5}$MnO$_3$ 
thin films grown on LSAT (011)$_c$ substrates. 
The different scattering angles together with the polarization 
and energy dependent lineshapes enable us 
to separate the orbital and magnetic scattering originating 
from different twin domains. 
In addition to the orbital order and spin order transitions known from 
bulk materials, we observe a third spin reorientation transition 
at $T_2$ that is absent in bulk materials. Importantly, the magnetic
(0,1/2,0)-modulation of the 
PCMO-film is different from the (0,1/2,1)-modulation of the conventional 
bulk CE-phase: in contrast to bulk materials, the film exhibits 
ferromagnetic correlations along the $c$-axis. We also deduce 
a different magnetic easy axis for the film as compared to the bulk. 
These differences between the PCMO-film and the corresponding 
bulk material indicate a direct coupling of the epitaxial 
strain to the magnetism. This coupling is most likely mediated 
by the orbital degree of freedom, which couples strongly 
to both the lattice and the spin system, thereby providing 
a convenient route to tune the magnetic properties of manganite films 
by epitaxial strain.  Furthermore, the presented analysis demonstrates 
that detailed information can be obtained from RSXS measurements already 
from arguments that are solely based on local symmetries. 
Even more information, however, could be extracted 
from a detailed lineshape analysis. 
The present study is the first demonstration of 
the antiferromagnetic ordering below 150 K 
in Pr$_{0.5}$Ca$_{0.5}$MnO$_3$ 
thin films. Since the volume of thin films 
is very small, it is difficult to study magnetism of thin films 
by neutron diffraction. Our results demonstrate that 
RSXS is a very suitable experimental technique to study 
magnetic order of thin film materials. 

H.~Wadati and J.~Geck contributed equally to this work. 
The authors would like to thank Y. Wakabayashi, U. Staub, 
A. Tanaka, and J. Okamoto for informative discussions. 
This research was made possible with financial support 
from the Canadian funding organizations NSERC, CFI, and 
CIFAR and is granted by the Japan Society for the Promotion of
Science (JSPS) through the ``Funding Program for 
World-Leading Innovative R\&D on Science and Technology 
(FIRST Program)'', initiated by the Council for Science 
and Technology Policy (CSTP). 
 J. Geck gratefully acknowledges the support through the 
DFG Emmy Noether Program (Grant GE-1647/2-1). 
\bibliography{LVO1tex}
\end{document}